\documentclass[a4paper,12pt,fleqn]{article}
\usepackage[DIV13]{typearea}
\usepackage{amsmath}
\usepackage{amssymb}
\usepackage{amsfonts}
\usepackage{amscd}
\usepackage{bbm} 
\usepackage[small]{caption2}
\usepackage{cite}
\usepackage{graphicx} 
\usepackage{mathrsfs}
\usepackage[center,footnotesize,hang]{subfigure}
\usepackage{hyperref}
\usepackage{url}

\newcommand{\CiteSeeSaw}{\cite{Minkowski:1977sc,Ramond:1979py,Yanagida:1980,Glashow:1979vf,Gell-Mann:1980vs,Mohapatra:1980ia}}

\newcommand{\eV}{\ensuremath{\,\mathrm{eV}}}

\newcommand{\nova}{NO$\nu$A}
\def\D{\mathrm{d}} 
\def\I{\mathrm{i}}

\DeclareMathOperator{\re}{Re}

\DeclareMathOperator{\Tr}{Tr}
\DeclareMathOperator{\diag}{diag}
\newcommand{\RaiseBrace}[1]{\raise1.5pt\hbox{$\displaystyle#1$}}

\newcommand{\package}[1]{\href{http://www.ph.tum.de/~rge/}{\tt #1}}

\allowdisplaybreaks[1]

\begin{document}

\begin{titlepage}

\ \vspace*{-15mm}
\begin{flushright}
TUM-HEP-589/05
\end{flushright}
\vspace*{5mm}

\begin{center}
{\Large\sffamily\bfseries 
Renormalization Group Evolution of}\\[3mm]
{\Large\sffamily\bfseries Dirac Neutrino Masses}
\\[13mm]
{\large
Manfred Lindner\footnote{E-mail: \texttt{lindner@ph.tum.de}}$^{(a)}$,
Michael Ratz\footnote{E-mail: \texttt{mratz@th.physik.uni-bonn.de}}$^{(b)}$ and
Michael Andreas Schmidt\footnote{E-mail: \texttt{mschmidt@ph.tum.de}}$^{(a)}$
}
\\[5mm]
{\small\textit{$^{(a)}$
Physik-Department T30, 
Technische Universit\"{a}t M\"{u}nchen\\ 
James-Franck-Stra{\ss}e,
85748 Garching, Germany
}}
\\[3mm]
{\small\textit{$^{(b)}$
Physikalisches Institut der Universit\"at Bonn,\\
Nussallee 12, 53115 Bonn, Germany.}}
\\[3mm]
\vspace*{2.0cm}
\end{center}
\begin{abstract}
\noindent
There are good reasons why neutrinos could be Majorana particles, but 
there exist also a number of very good reasons why neutrinos could 
have Dirac masses. The latter option deserves more attention and we 
derive therefore analytic expressions describing 
the renormalization group evolution of mixing angles and of the 
CP phase for Dirac neutrinos. Radiative corrections to leptonic mixings 
are in this case enhanced compared to the quark mixings because the 
hierarchy of neutrino masses is milder and because the mixing angles 
are larger. The renormalization group effects are compared to the 
precision of current and future neutrino experiments. We find that, in the 
MSSM framework, radiative corrections of the mixing angles are 
for large $\tan\beta$ comparable to the precision of future experiments. 
\end{abstract}

\end{titlepage}

\newpage
\setcounter{footnote}{0}

\section{Introduction}

One of the most important open questions of neutrino physics is  whether
neutrinos are Dirac or Majorana particles. From a theoretical perspective, large
Majorana mass terms appear quite naturally for  the right-handed neutrinos,
since they are complete gauge singlets.  This leads directly to the best
investigated (and therefore most  popular) explanations for the huge ratio of
observed mass scales  in the see-saw mechanism~\CiteSeeSaw. In its simplest
form, neutrino masses get suppressed by a factor $v_\mathrm{EW}/M_*$ with
$v_\mathrm{EW}$ denoting the vacuum expectation value (VEV) of the Higgs boson
and $M_*$ being the scale at which $B-L$ symmetry (baryon minus lepton number)
is assumed to be broken. However, it is important to keep in mind that the
suppression factor  $v/M_*$ (with $M_*$ now being the GUT scale or a related
scale) arises  rather generally whenever neutrino masses arise from integrating
out  heavy degrees of freedom with mass $M_*$. This statement holds
independently  of the nature of neutrino masses, in particular both for Dirac
and Majorana  masses. Indeed, there exist a couple of appealing models where
small  Dirac masses are explained in this way by using extra, heavy degrees of
freedom, or by relating the Yukawa coupling $Y_\nu$ to the ratio of  gravitino
mass (or other soft masses) and GUT (or compactification)
scale~\cite{Mohapatra:1986bd,Arkani-Hamed:2000bq,Borzumati:2000mc,Kitano:2002px,Arnowitt:2003kc,Abel:2004tt}.
Another possibility is using localization in extra dimensions, and  explaining
the suppression by a small overlap of the corresponding  zero-mode profiles
along extra  dimensions (see, e.g., \cite{Hung:2004ac,Ko:2005sh,Antusch:2005kf}).  Further
support for Dirac neutrinos was found in certain orbifold  compactifications of
the heterotic string, where it is difficult to  obtain the standard see-saw
\cite{Giedt:2005vx}. For recent overviews and further references of various
possibilities of explaining small Dirac masses see
\cite{Murayama:2004me,Smirnov:2004hs,Mohapatra:2004vr}. 

Cosmological arguments do not give a preference for Dirac or Majorana 
masses either. For instance, even if one requires the explanation of 
the observed baryon asymmetry to be related to neutrino properties, 
one finds that successful baryogenesis can work both for 
Majorana \cite{Fukugita:1986hr,Buchmuller:2005eh} and 
Dirac \cite{Dick:1999je,Murayama:2002je} neutrinos. Dirac neutrinos 
evade also constraints from primordial nucleosynthesis, since the 
right-handed degrees of freedom decouple with a low temperature so 
that their energy density is relatively suppressed \cite{Dolgov:1994vu}. 
The question whether neutrinos are Dirac or Majorana particles is 
therefore one of the main motivations for improved neutrino-less 
double beta decay experiments. Both options should therefore be 
studied seriously until this question is clarified by experiments.

We investigate in this Letter RG effects under the assumption that 
neutrinos are Dirac particles, and that the small Yukawa couplings 
are explained by some mechanism operating at a high, e.g.\ GUT or 
compactification, scale, denoted by $M_\mathrm{GUT}$ in the following. 
The radiative corrections to the leptonic CP violation has been studied in \cite{Chiang:2000um}. We extend this analysis to include all leptonic mixing parameters, and derive analytic formulae describing the renormalization group (RG) running of the leptonic mixing parameters. The radiative 
corrections are compared to analogous corrections in the quark sector, 
and we will show that generically RG running in the lepton sector is 
stronger than in the quark sector since the coefficients of the 
renormalization group equations (RGEs) are enhanced due to the fact 
that mass hierarchy is milder and the mixing angles are larger in the 
lepton sector. We compare the size of the radiative corrections to 
the accuracy of present and future neutrino experiments, and find that 
they are particularly relevant if neutrino masses are degenerate 
and $\tan\beta$ is large in the MSSM.

\section{Analytic formulae}
\label{sec:AnalyticFormulae}

Using the standard parameterization (see, e.g., \cite{Eidelman:2004wy}) 
for leptonic (and quark) mixing the RGEs for the leptonic mixing angles
read
\begin{eqnarray}
 \Dot{\theta}_{12}
 & = & \frac{-C\,y_\tau ^2}{32\,\pi^2}\,
        \frac{m_1^2 +m_2^2}{m_2^2 -m_1^2}\,
        \sin (2\,\theta_{12})\,\sin^2 \theta_{23}
 +\mathscr{O}(\theta_{13})\;,
 \label{eq:DotTheta12}\\
 \Dot{\theta}_{13}
 & = & 
 \frac{-C\,y_\tau^2}{32\,\pi^2}\frac{1}{\left( m_3^2 - m_1^2 \right) \,
    \left( m_3^2-m_2^2 \right)}
 \left\{  \left(  m_2^2 - m_1^2  \right) \,
       m_3^2\,\cos \delta\,\cos \theta_{13}\,
       \sin (2\,\theta_{12})\,\sin (2\,\theta_{23})
 \right.\nonumber\\
 & & \left.\quad{}
  +\left[  m_3^4 -  \left( m_2^2 - m_1^2 \right) \,
            m_3^2\,\cos (2\,\theta_{12}) - m_1^2\,m_2^2 \right] \,
         \cos^2\theta_{23}\,\sin (2\,\theta_{13})
   \right\}\;,
 \label{eq:DotTheta13}\\
 \Dot{\theta}_{23}
 & = &
 \frac{-C\,y_\tau^2}{32\,\pi^2}\,
 \frac{\left[ m_3^4-m_1^2\,m_2^2  + ( m_2^2 - m_1^2 ) \,m_3^2\,
         \cos (2\,\theta_{12}) \right]}{
                ( m_3^2 -m_1^2) \,( m_3^2-m_2^2 ) }\,
 \sin (2\,\theta_{23})
 +\mathscr{O}(\theta_{13})\;,\label{eq:DotTheta23}
\end{eqnarray}
where the dot indicates the logarithmic derivative w.r.t.\ the renormalization
scale $\mu$, e.g.\ $\Dot{\theta}_{12}=\D\theta_{12}/\D t
=\mu\,\D\theta_{12}/\D\mu$, and
\begin{equation}
 C~=\left\{\begin{array}{ll}
 1\;,\quad& \text{(MSSM)}\;,\\
 -3/2\;,\quad& \text{(SM)}\;.
 \end{array}\right.
\end{equation}
Here, we have neglected the tiny electron and muon Yukawa couplings, 
as well as the neutrino Yukawa couplings, against the $\tau$ Yukawa 
coupling $y_\tau$. Furthermore, in $\Dot{\theta}_{12}$ and 
$\Dot{\theta}_{23}$ we only kept the leading order term of an expansion 
in the reactor mixing angle $\theta_{13}$. We have checked numerically 
that this is a good approximation for realistic values of $\theta_{13}$.

It is instructive to compare Eqs.~\eqref{eq:DotTheta12}--\eqref{eq:DotTheta23}
to the corresponding expressions for Majorana neutrinos. Technically 
one obtains Eqs.~\eqref{eq:DotTheta12}--\eqref{eq:DotTheta23} by 
discarding all terms which depend on the Majorana phases in 
Eqs.~(8)--(10) of  Ref.~\cite{Antusch:2003kp}. One could thus say 
that the running of the Dirac mixing parameters resembles the
running of Majorana mixing parameters averaged over the Majorana phases
$\varphi_1$ and $\varphi_2$\footnote{Stated differently, the running in the
Dirac case is roughly half of the maximal running in the Majorana case. The
factor 1/2 can be understood from the structure of the RGE: in the Dirac case,
the mass matrix gets rotated by only one term (cf.\ Eq.~\eqref{eq:DiracRGE}),
\[\Delta m_\nu~=~ C\,m_\nu\,\left(Y_e^\dagger Y_e\right)+\text{flavor-trivial
terms}\;,\] 
while in the Majorana case there are two terms
\[\Delta m_\nu~=~ C\,\left[m_\nu\,\left(Y_e^\dagger Y_e\right)
+\left(Y_e^\dagger Y_e\right)^T m_\nu\right]+\text{flavor-trivial
terms}\;,\] 
with $C=-3/2$ in SM \cite{Antusch:2001ck} and two-Higgs models
\cite{Antusch:2001vn}, and $C=1$ in the MSSM
\cite{Chankowski:1993tx,Babu:1993qv}.}. This means in particular that strong
damping effects for the evolution of $\theta_{12}$, as observed for Majorana
neutrinos, cannot occur in the Dirac case.

From Eqs.~\eqref{eq:DotTheta12}--\eqref{eq:DotTheta23}, we can immediately
recognize several features of the RG evolution. First, for a strong mass
hierarchy, the running of the angles is negligible. For $m_1=0$, the angles
always run less than $1^\circ$  (except for $\theta_{23}$  which runs more if
$\tan\beta\gtrsim 40$).  Second, for $m_1\gtrsim 0.02\,\mathrm{eV}$,
$\theta_{12}$ has the strongest RG  evolution. Third, as is obvious from 
Eqs.~\eqref{eq:DotTheta12} and \eqref{eq:DotTheta23}, in the MSSM $\theta_{12}$
always increases when running downwards while $\theta_{23}$ increases for a
normal and decreases for an inverted mass hierarchy. 
This means that these two angles are radiatively enhanced (for normal 
mass ordering) which may, at least partially, be the reason for their 
large size. Whether $\theta_{13}$ increases or decreases depends on $\delta$.

The evolution of the Dirac phase $\delta$ is described by\footnote{The evolution of the weak basis CP invariant has already been studied in~\cite{Chiang:2000um}.}
\begin{equation}\label{eq:DotDelta}
 \Dot{\delta}
 ~=~
 \Dot{\delta}^{(-1)} \theta_{13}^{-1}
 +
 \Dot{\delta}^{(0)} 
 +\Dot{\delta}^{(1)} 
 +
 \mathscr{O}\bigl(\theta_{13}^2\bigr)\;,
\end{equation}
with the first two coefficients $\Dot{\delta}^{(k)}$ given by 
\begin{subequations}
\begin{eqnarray}
 \Dot{\delta}^{(-1)} 
 & = &  
 \frac{C\,y_\tau ^2}{32\,\pi^2}\,
 \frac{(m_2^2-m_1^2)\,m_3^2}{\left( m_3^2 - m_1^2 \right) \,
    \left( m_3^2 - m_2^2 \right)}
 \,\sin (\delta )\,\sin (2\,{{\theta }_{12}})\,\sin (2\,{{\theta }_{23}})
 \;, \\
 \Dot{\delta}^{(0)} 
 & = & 0\;.
\end{eqnarray}
Moreover, the term linear in $\theta_{13}$ contains
\begin{eqnarray}
 \Dot{\delta}^{(1)} 
 & = &  
 \frac{C\, y_\tau^2}{16\,\pi^2}\,
 \frac{m_2^2\,\left( m_3^2 - m_1^2 \right)^2}{%
        \left( m_2^2 - m_1^2 \right) \,\left( m_3^2 - m_1^2 \right) \,
    \left( m_3^2 - m_2^2 \right) }
 \cot ({{\theta }_{12}})\,\sin (2\,{{\theta }_{23}}) \,\sin\delta
 +\dots\;,
\end{eqnarray}
\end{subequations}
which becomes relevant if $\theta_{13}$ is not too small.

\begin{figure}[ht]
 \centerline{%
 \subfigure[$\sin^2(2\theta_{13})$.]{\includegraphics[scale=0.8]{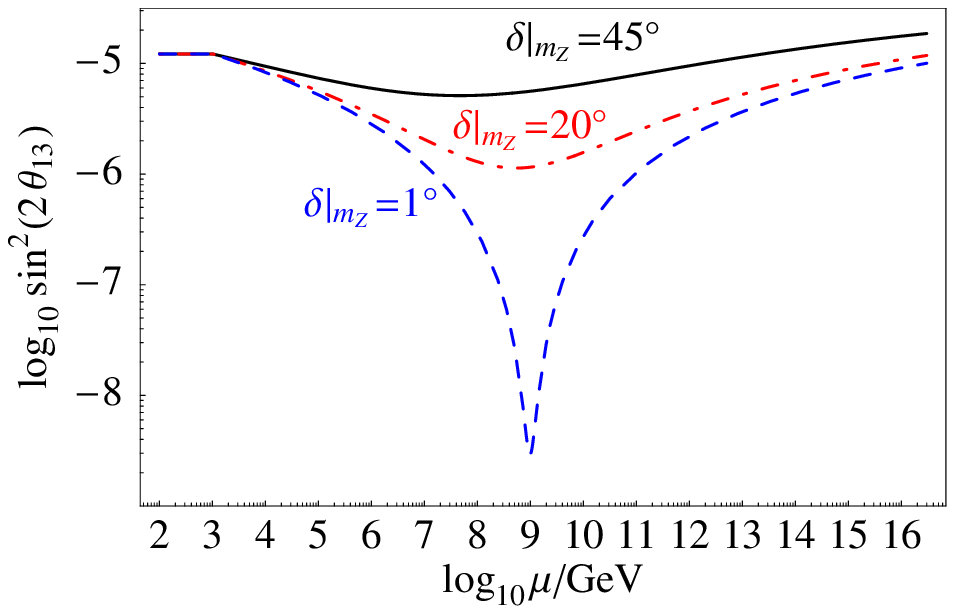}}
 \quad
 \subfigure[$\delta$.]{\includegraphics[scale=0.8]{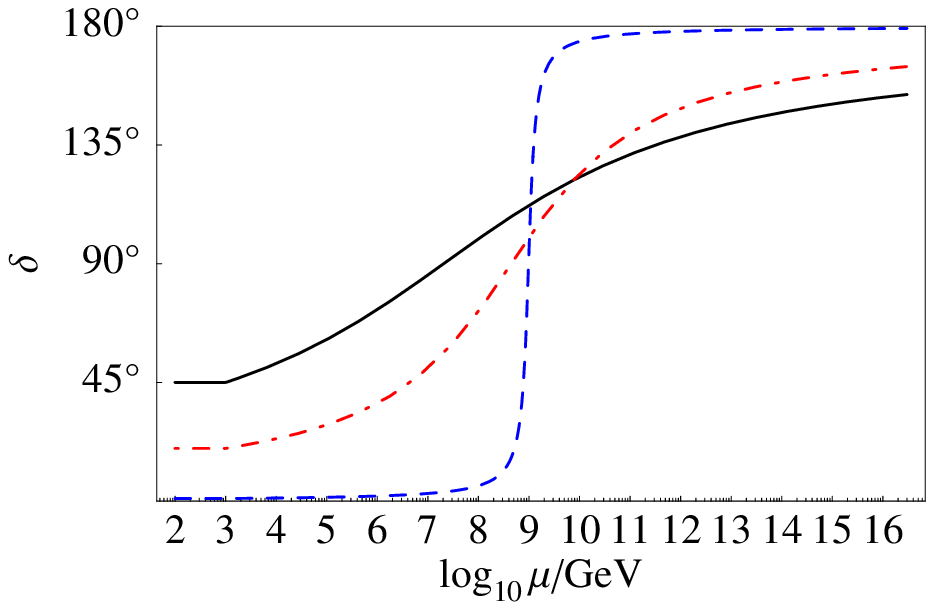}}
 }
\caption{Evolution of (a) $\sin^2(2\theta_{13})$ and (b) $\delta$ for small
 values of $\theta_{13}$. We use $\tan\beta=50$, $\theta_{13}=0.1^\circ$,
 $m_1=0.1\,\mathrm{eV}$ and best-fit values for all other parameters. The solid
 (black), dash-dotted (red) and dashed (blue) curve shows the evolution of (a)
 $\theta_{13}$ and (b) $\delta$ for $\delta=45^\circ$, $20^\circ$ and $1^\circ$
 at the electroweak scale, respectively.  $\theta_{13}$ cannot become 0 in any
 of the cases.}
\label{fig:Theta13CloseTo0}
\end{figure}

As usual, $\delta$ is undefined for $\theta_{13}=0$. Clearly, if $\delta$
vanishes for some scale, the (lepton sector of the) theory is CP invariant
at this scale and thus remains CP invariant for all scales. Hence, the 
statement $\delta=0$ cannot depend on the renormalization scale, which
can also be seen in our formulae. Likewise, if $\theta_{13}$ is zero at 
some given scale, the theory must again be CP invariant for all 
scales\footnote{This is in contrast to the case of Majorana neutrinos, 
where the memory to CP violation can be stored in the Majorana phases, 
and $\theta_{13}$ can cross zero even in the presence of leptonic CP 
violation \cite{Casas:1999tg,Antusch:2003kp}.}. From this we conclude 
that $\theta_{13}$ can never cross zero if we have at some scale
$\theta_{13}\ne0$ and $\sin\delta\ne 0$. If $\theta_{13}$ approaches 
zero, then we can see from \eqref{eq:DotDelta} that $\delta$ runs 
quickly to a value such that the coefficient in \eqref{eq:DotTheta13} 
changes its sign and $\theta_{13}$ increases again. Thus, the only 
case where $\theta_{13}$ can cross zero is the CP conserving case.
This is interesting for future precision measurements of $\theta_{13}$,
since the assumption of leptonic CP violation at any scale leads 
to the conclusion that the weak-scale value of $\theta_{13}$ does 
not vanish. We illustrate the corresponding large effects in the 
evolution of $\delta$ in Fig.~\ref{fig:Theta13CloseTo0}. 
For all our plots, we use the software packages \package{REAP} and 
\package{MPT} associated with \cite{Antusch:2005gp}.

We can understand this feature also differently. In the above 
approximation, we can write $U_{e3} = \theta_{13}e^{-\I\delta}$ 
and by inserting the RGEs for $\theta_{13}$ and $\delta$, we find 
in the limit $\theta_{13}\to0$
\begin{eqnarray}
 \frac{\D}{\D t}U_{e3}
 & = & 
 \Dot{\theta}_{13}\,e^{-\I\delta}-\I\,\theta_{13}\,e^{-\I\delta}\,\Dot{\delta}
 \nonumber\\
 & \simeq &
 \frac{C\,y_\tau ^2}{32\,\pi^2}\,
 \frac{(m_2^2-m_1^2)\,m_3^2}{\left( m_3^2 - m_1^2 \right) \,
    \left( m_3^2 - m_2^2 \right)}
 \,\sin (2\,{{\theta }_{12}})\,\sin (2\,{{\theta }_{23}})~.
\label{eq:Ue3Dot}
\end{eqnarray}
For $\theta_{13}\to0$ we find thus that the RG change of $U_{e3}$ 
is along the real axis and $U_{e3}$ can therefore only become zero 
if it is real. The imaginary part of Eq.~\eqref{eq:Ue3Dot} allows 
to determine a minimal value of $\theta_{13}$ as
$(\theta_{13})_\mathrm{min}\simeq \theta_{13}(\mu)\,\sin\delta(\mu)$ 
where any scale $\mu$ can be used.

Furthermore, let $t_0$ denote the turning point of $\theta_{13}$ characterized
by $\delta=\pi/2$. The `asymptotic' behavior 
$\delta(t-t_0)\simeq-\delta(t_0-t)=\pi-\delta(t_0-t)$ is a consequence of the
fact that $\Dot{\delta}$ is an odd function in $\theta_{13}$\footnote{We have
introduced $\pi$ in order to keep $\theta_{13}$ positive as we use the
convention that $\theta_{13}$ is always positive, and a possible sign flip is
absorbed in a jump of $\delta$.}.  This allows to understand why $\delta$
approaches $\pi-\delta(m_Z)$ for large $\mu$ in
Fig.~\ref{fig:Theta13CloseTo0}.

The evolution of the mass eigenvalues is given by
\begin{subequations}
\begin{eqnarray}
 16\pi^2\,\Dot{m}_1
 & = & 
 \left\{
 C\,y_\tau^2
 \left[ \cos^2 \theta _{12}\,
       \cos^2 \theta_{23}\,\sin^2 \theta_{13} + 
      \sin^2 \theta_{12}\,\sin^2 \theta_{23} 
 \vphantom{\frac{1}{2}}\right.\right.
 \nonumber\\
 & & \hphantom{ \left\{ y_\tau^2 \left[\right.\right.}
 \left.\left.
      {}-\frac{1}{2}\cos \delta\,\sin \theta_{13}\,
          \sin (2\,\theta_{12})\,
      \sin (2\,\theta_{23})
 \right]
          +\alpha_\nu
 \right\}\,m_1
 \;,\\
 16\pi^2\,\Dot{m}_2
 & = &             
 \left\{ C\,y_\tau^2
 \left[ \sin^2 \theta_{12}\,\cos^2 \theta_{23}\,\sin^2 \theta_{13} + 
      \cos^2 \theta_{12}\,\sin^2 \theta_{23} 
 \vphantom{\frac{1}{2}}\right.\right.
 \nonumber\\
 & & \hphantom{ \left\{ y_\tau^2 \left[\right.\right.}
 \left.\left.
          {}+\frac{1}{2}\cos \delta\,\sin \theta_{13}\,\sin (2\,\theta_{12})\,
       \sin (2\,\theta_{23}) 
 \right]
          +\alpha_\nu
 \right\}\,m_2
 \;,\\
 16\pi^2\,\Dot{m}_3
 & = & 
 \left\{ C\,y_\tau^2\,\cos^2\theta_{13}\,\cos^2 \theta_{23}
 +\alpha_\nu
 \right\}\,m_3
 \;.
\end{eqnarray}
\end{subequations}
$\alpha_\nu$ represents the flavor-independent part of the RGE, and is given in
\eqref{eq:AlphaNu}. Clearly, the dominant RG effect of the masses is a common
rescaling governed by $\alpha_\nu$. In addition, for large $\tan\beta$ in the
MSSM, there are corrections specific to the individual $m_i$. In leading order
in $\theta_{13}$, these effects tend to decrease $\Delta m^2_\mathrm{atm}$ for
a normal hierarchy and to increase $\Delta m^2_\mathrm{atm}$ for an inverted
hierarchy when running down.

\begin{figure}[ht]
 \centerline{%
 \subfigure[$m_i(\mu)/m_i(m_Z)$.]{%
        \includegraphics[scale=0.86]{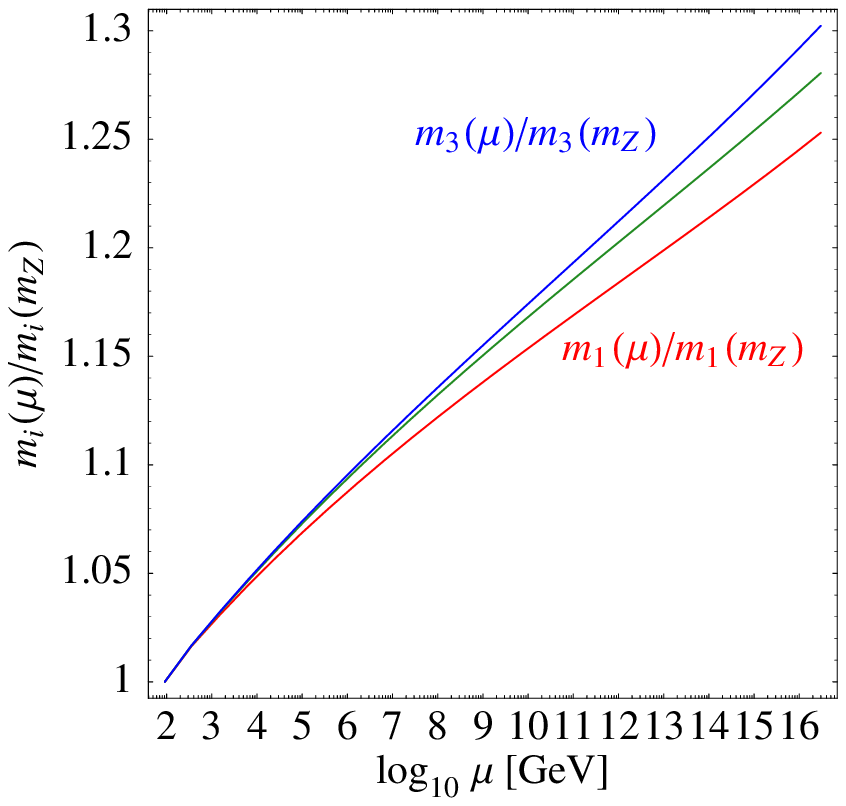}}
 \quad
 \subfigure[$\Delta m^2_{ij}(\mu)/\Delta m^2_{ij}(m_Z)$.]{%
        \includegraphics[scale=0.85]{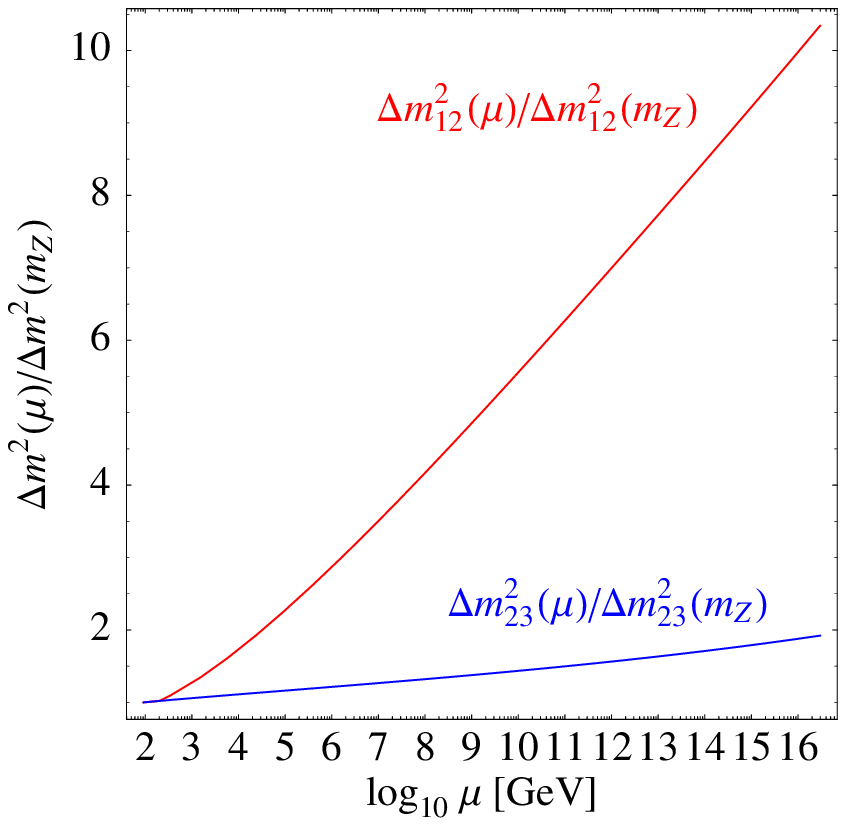}}
 }
\caption{Example of the evolution of the mass eigenvalues and the 
$\Delta m^2$s. We choose $m_1(m_Z)=0.1\,\mathrm{eV}$, 
$\delta(m_Z)=50^\circ$, $\tan\beta=50$ and a SUSY breaking scale 
of $200\,\mathrm{GeV}$, and best-fit values otherwise.}
\label{fig:ScalingOfMassesDirac}
\end{figure}

Fig.~\ref{fig:ScalingOfMassesDirac} shows an extreme example of the 
evolution of the mass eigenvalues and the corresponding 
$\Delta m^2_\mathrm{sol}$ and $\Delta m^2_\mathrm{atm}$. For 
large $\tan\beta$, there are substantial deviations from
the flavor-independent scaling of the mass eigenvalues. The latter 
can be approximately inferred from the curve of $m_1$ in
Fig.~\ref{fig:ScalingOfMassesDirac} (a).

\section{Radiative corrections and future precision experiments}

An important question is if future experiments will reach a  precision which
allows to draw interesting conclusions from quantum corrections. There exist,
for example, models where  $\theta_{13}$ vanishes at the GUT scale, but RG
corrections  still lead to a finite value of $\theta_{13}$ at  low energies. A
certain finite value of $\theta_{13}$ is  therefore guaranteed unless the
initial values at the GUT  scale and the independent RG effects are adjusted to
cancel  each other. From the discussion of the previous section, we would also
know that the CP phase $\delta$ would vanish for all scales for Dirac neutrinos,
while it could become finite for Majorana neutrinos. A finite value of 
$\delta$ and $\theta_{13}$ would thus exclude either Dirac neutrinos or 
$\theta_{13}(M_{GUT})=0$. Similar arguments can be
made for other quantities of the leptonic mixing matrix. $\theta_{23}$, for
example, is within  current experimental errors compatible with $45^\circ$. 
However, as in the case of $\theta_{13}$, certain deviations  are expected from
RG effects even if $45^\circ$ is exactly  predicted at the GUT scale. Future
precision measurements  of neutrino oscillations may therefore allow interesting
tests  of flavor models and related renormalization group effects. 

Atmospheric neutrino data~\cite{Suzuki:2004pb} and results from 
the K2K long-baseline accelerator  experiment~\cite{Suzuki:2004pb}
currently determine 
$\Delta m_{31}^2 = (2.2^{+0.6}_{-0.4})\times10^{-3}\eV^{2}$ and 
$\theta_{23}\approx45^{\circ}$~\cite{Suzuki:2004pb,Gonzalez-Garcia:2004jd}, 
whereas solar neutrino
data~\cite{Fukuda:2002pe,Cleveland:1998nv,Abdurashitov:1999zd,Hampel:1998xg,Cattadori:2002rd,Ahmad:2002jz,Ahmad:2002ka,Ahmed:2003kj},
combined with  the results from the KamLAND reactor
experiment~\cite{Suzuki:2004pd} lead to $\Delta m_{21}^{2} =
(8.2^{+0.3}_{-0.3}) \times 10^{-5}\eV^2$  and $\tan^2\theta_{12} =
0.39^{+0.05}_{-0.04}$~\cite{Gonzalez-Garcia:2004jd}. These results are 
to a good approximation still described by two independent two flavor 
oscillations. The key parameter for genuine three flavor effects is 
the  mixing angle $\theta_{13}$ which is so far only known to be small  
from the CHOOZ~\cite{Apollonio:1997xe,Apollonio:2002gd} and Palo
Verde~\cite{Boehm:1999gk} experiments. The current bound for $\theta_{13}$ 
depends on the value of the atmospheric mass squared difference and it 
gets weaker for $\Delta m^2_{31} \lesssim 2\times 10^{-3}\eV^2$. However, 
in that region an additional constraint on $\theta_{13}$ from global 
solar neutrino  data becomes important~\cite{Maltoni:2004ei,Maltoni:2003da}. 
At the current best fit value of $\Delta m^2_{31} = 2.2\times 10^{-3}\eV^2$ 
the $3\sigma$ bound is
$\sin^2\theta_{13} \le 0.041$~\cite{Gonzalez-Garcia:2004jd}. 

Genuine three flavor oscillation effects occur only for a finite value 
of $\theta_{13}$ and establishing a finite value of $\theta_{13}$ is 
therefore one of the next milestones in neutrino physics. Leptonic CP 
violation is another three flavor effect which can only be tested 
if $\theta_{13}$ is finite. There exists therefore a very strong 
motivation to establish a finite value of $\theta_{13}$ and then leptonic CP 
violation~\cite{Freund:2000ti,Freund:2001ui,Huber:2002rs,Lindner:2002vt,Huber:2002mx}.
Different experimental projects are therefore under construction 
or are being planned in order to achieve these goals. 
It is useful to divide the future into what can be achieved with 
specific current or next generation projects and what may be achieved 
with long term projects.
The MINOS~\cite{Ables:1995wq} project, which started already data 
taking, and the CNGS projects ICARUS~\cite{Aprili:2002wx} and 
OPERA~\cite{Duchesneau:2002yq}, which are completing construction
can be considered as ``current projects''. Beyond that exist other, 
more ambitious ``next generation'' projects like the superbeam 
experiments J-PARC to SuperKamiokande (T2K)~\cite{Itow:2001ee} and 
the NuMI off-axis experiment \nova~\cite{Ayres:2002nm}. In 
addition there are ``next generation'' plans for new reactor neutrino 
experiments~\cite{Anderson:2004pk} with a near and far detector. 
A first interesting question concerns improvements of $\Delta m^2_{31}$ 
and $\sin^2\theta_{23}$. In Tab.~\ref{tab:atm} we show the relative 
precision which can be obtained in the future 
in comparison to the current precision, as obtained from a global 
fit to SuperKamiokande (SK) atmospheric and K2K long-baseline 
data~\cite{Maltoni:2004ei,Maltoni:2003da}. We observe from these 
numbers, that the accuracy on $\Delta m^2_{31}$ can be improved 
by one order of magnitude, whereas the accuracy on 
$\sin^2\theta_{23}$ will be improved only by a factor two.

\begin{table}[htb]
\centering
\begin{tabular}{lrr}
\hline
 & $|\Delta m^2_{31}|$  & $\sin^2\theta_{23}$ \\
\hline
current     & 88\% & 79\% \\
\hline
MINOS+CNGS  & 26\% & 78\% \\
T2K         & 12\% & 46\% \\
\nova\        & 25\% & 86\% \\
\hline
Combined &  9\% & 42\% \\
\hline
\end{tabular}
\caption{Relative precision of $|\Delta m^2_{31}|$ and 
  $\sin^2\theta_{23}$ at $3\sigma$ for the values 
  $\Delta m^2_{31} = 2\times 10^{-3}\eV^2$,
  $\sin^2\theta_{23} = 0.5$. The last row is the 
  relative precision which can be obtained by combining 
  all experiments (from \cite{Huber:2004dv}). 
}
\label{tab:atm}
\end{table}

Tab.~\ref{tab:atm} depends on the value of $\Delta m^2_{31}$ and
the sensitivity suffers for all experiments for low values of 
$\Delta m^2_{31}$. T2K will provide a relatively precise
determination of $\Delta m^2_{31}$ 
for $\Delta m^2_{31} \gtrsim 2\times 10^{-3}\eV^2$. Although \nova\ 
can put a comparable lower bound on $\Delta m^2_{31}$, the upper bound is
significantly weaker, and similar to the bound from MINOS
\cite{Huber:2004dv}. The reason for this
is a strong correlation between $\Delta m^2_{31}$ and $\theta_{23}$, which
disappears only for $\Delta m^2_{31} \gtrsim 3\times 10^{-3}\eV^2$.
From Tab.~\ref{tab:atm} one can also see that only T2K is able to 
improve the current bound on $\sin^2\theta_{23}$. The main reason 
for the rather poor performance on $\sin^2\theta_{23}$ is the fact 
that these experiments are mostly sensitive to $\sin^22\theta_{23}$. 
This implies that for $\theta_{23} \approx \pi/4$ it is very hard 
to achieve a good accuracy on $\sin^2\theta_{23}$, although
$\sin^22\theta_{23}$ can be measured with relatively high
precision~\cite{Minakata:2004pg}.

An important task of the next generation long baseline and reactor
experiments of the coming years will be to establish the three 
flavored-ness of the oscillations. The sensitivity 
to a finite value of the key parameter $\theta_{13}$ is only modest 
for MINOS, OPERA and ICARUS. Double Chooz, T2K and \nova\ can 
do much better. The $\sin^22\theta_{13}$-limits of the beam 
experiments are, however, strongly affected by parameter correlations 
and degeneracies, whereas new reactor experiments provide a ``clean'' 
measurement of $\sin^22\theta_{13}$~\cite{Huber:2003pm}. Altogether 
these experiments will provide an improvement by about a factor ten 
for $\sin^2 2\theta_{13}$ over the existing limit. In addition, the 
KamLAND \cite{Suzuki:2004pd} (and solar neutrino) data will 
also further increase the accuracy for $\Delta m^2_{21}$ and 
$\theta_{12}$. An accuracy of $5\%$ for $\Delta m_{12}^{2}$ and
$20\%$ for $\sin^2\theta_{12}$ is expected.
Further improvements are possible, e.g. by loading the 
SuperKamiokande detector with Gadolinium, which might lead to
an error of $3\%$  for $\Delta m_{21}^{2}$ and $15\%$ for 
$\sin^2\theta_{12}$, both at 99\%CL \cite{Choubey:2004bf}.

Beyond the next generation accelerator and reactor based 
oscillation experiments exist much more ambitious projects like 
the JHF-HyperKamiokande project, beta beams and neutrino 
factories\footnote{See \cite{Huber:2005jk} for a comparison and 
for references.}. 
Such experiments clearly require further R\&D before they can be built. 
However, assuming current knowledge, such facilities appear to be
possible and they will lead to a precision at the level of percent 
or even below. With a neutrino factory, for example, a sensitivity 
to a finite value of $\sin^22\theta_{13}$  might be achievable 
down to $10^{-5}$.

It is interesting to compare these perspectives with RG effects. 
To illustrate the RG effects, we start with initial values for the mixing parameters at the GUT scale, $M_\mathrm{GUT}=3\times10^{16}\,\mathrm{GeV}$, assuming that these values find an explanation in a more fundamental theory.\footnote{One could, for instance, enjoy the possibility of fixing the initial values by continuous (e.g.~\cite{Gogoladze:2001kj}) or discrete (e.g.~\cite{Hagedorn:2005kz}) symmetries. In this case, RG effects add to the corrections arising from the breakdown of those symmetries.} These initial values are then compared with the corresponding mixing parameters at $m_Z$. In all our illustrations, we assume  $m_\mathrm{SUSY}=1\,\mathrm{TeV}$,
and a normal mass hierarchy.  The simple expressions
(Eqs.~\eqref{eq:DotTheta12}--\eqref{eq:DotTheta23} and \eqref{eq:DotDelta})  allow a
quick estimate of the RG effects. A more precise evaluation requires a numerical
analysis for which we use the Mathematica package   \package{REAP}
\cite{Antusch:2005gp}, which is publicly  available\footnote{See
\href{http://www.ph.tum.de/~rge/}{\url{http://www.ph.tum.de/~rge/}}}.

\begin{figure}[!ht]
 \centerline{%
  \subfigure[$\Delta_\mathrm{RG}\theta_{12}\equiv
   \theta_{12}(m_Z)-\theta_{12}(M_\mathrm{GUT})$.]{%
        \includegraphics[scale=0.85]{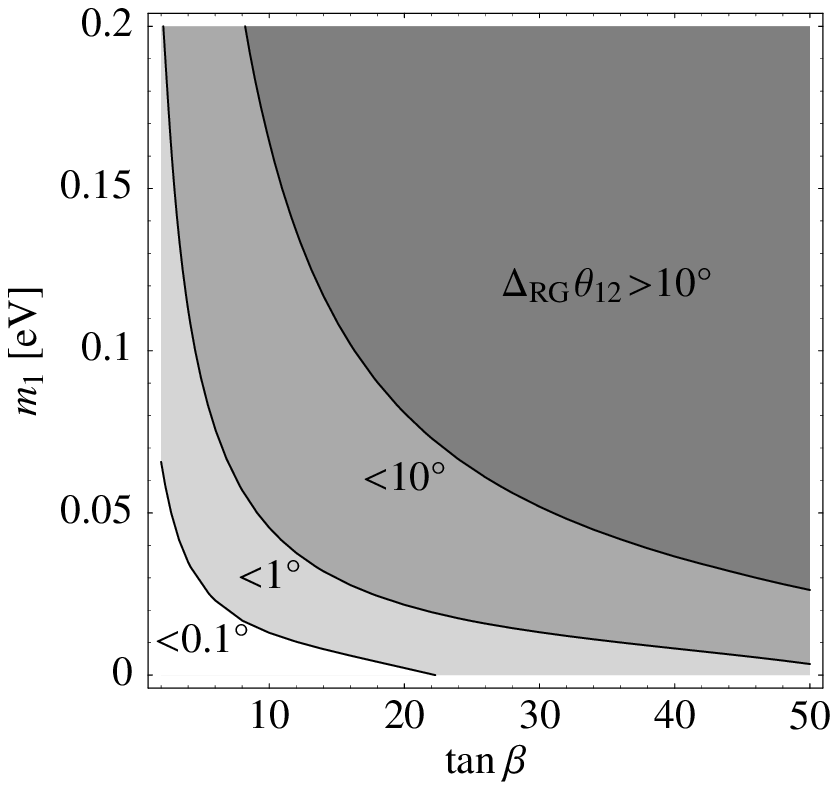}}
  \quad
  \subfigure[$\Delta_\mathrm{RG}\theta_{23}\equiv
   |0.5-\sin^2\theta_{23}(M_\mathrm{GUT})|$.]{%
        \includegraphics[scale=0.85]{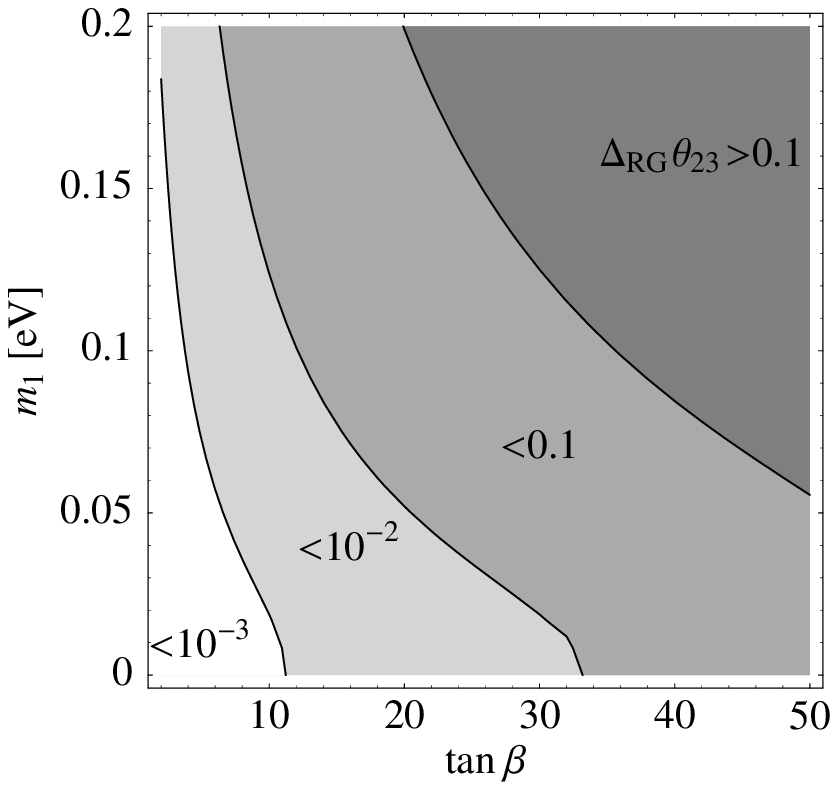}}
 }
 \caption{Radiative correction to (a) $\theta_{12}$ and (b) $\theta_{23}$ for
  $\theta_{12}=33^\circ$, $\theta_{13}=10^\circ$, $\theta_{23}=45^\circ$ and
  $\delta=90^\circ$ at $\mu=m_Z$ as a function of $\tan\beta$ and $m_1$. These
  contours remain to a large extent unchanged when varying $\theta_{13}$ in the
  allowed range and $\delta$ arbitrarily.}
\label{fig:DeltaRGTheta12and23}  
\end{figure}

The mixing angles $\theta_{12}$ and $\theta_{23}$ turn out to be rather unstable
for a degenerate spectrum  (cf.\ Fig.~\ref{fig:DeltaRGTheta12and23}). As a
consequence, a Dirac version of quark-lepton complementarity
\cite{Raidal:2004iw,Minakata:2004xt,Frampton:2004vw} can only work  for certain
mass eigenvalues and ratios of the Higgs VEVs and  $\tan\beta$ (for the
discussion of the RG effects in the see-saw Majorana case see
\cite{Antusch:2005gp,Mei:2005qp,Ellis:2005dr}). This means stability of
$\theta_{12}$ is only given  in models with hierarchical masses and/or small
$\tan\beta$.  Note also that for an inverted hierarchy $\theta_{12}$ is
unstable. This means that concerning $\theta_{12}$ RG effects are in general an
issue. 
RG corrections to the special value $\theta_{23}=45^\circ$ can 
be comparable to the precision of upcoming experiments. Again, 
this happens for a quite degenerate spectrum and/or large 
$\tan\beta$. 

\begin{figure}[!h]
 \centerline{%
  \subfigure[$\theta_{13}(m_Z)=10^\circ$;
  $\delta(m_Z)=90^\circ$.]{%
        \includegraphics[scale=0.85]{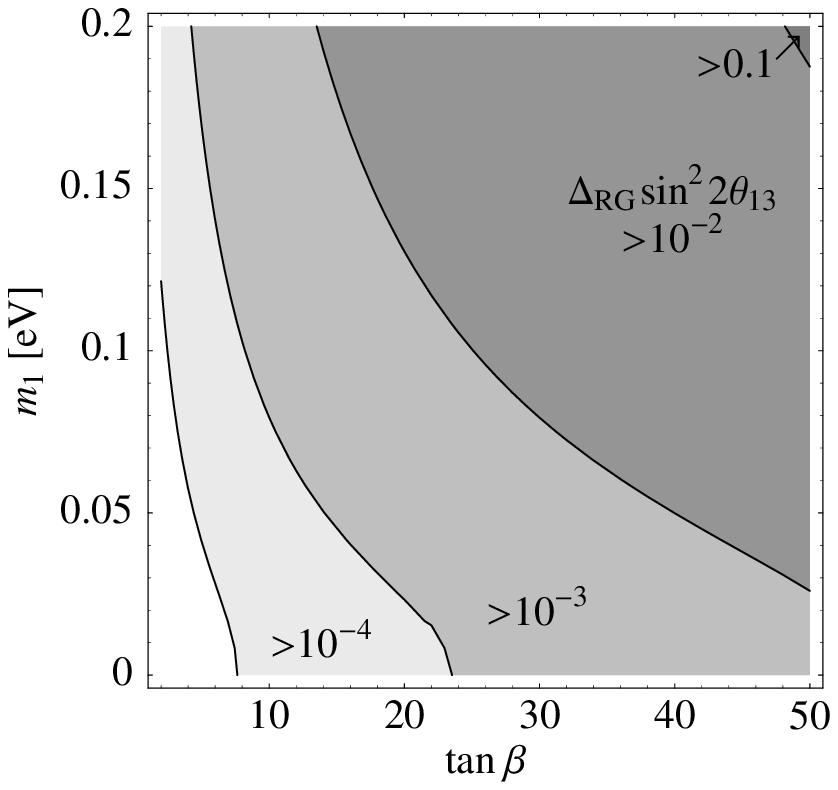}}
 \quad
  \subfigure[$\theta_{13}(m_Z)=0$.]{%
        \includegraphics[scale=0.85]{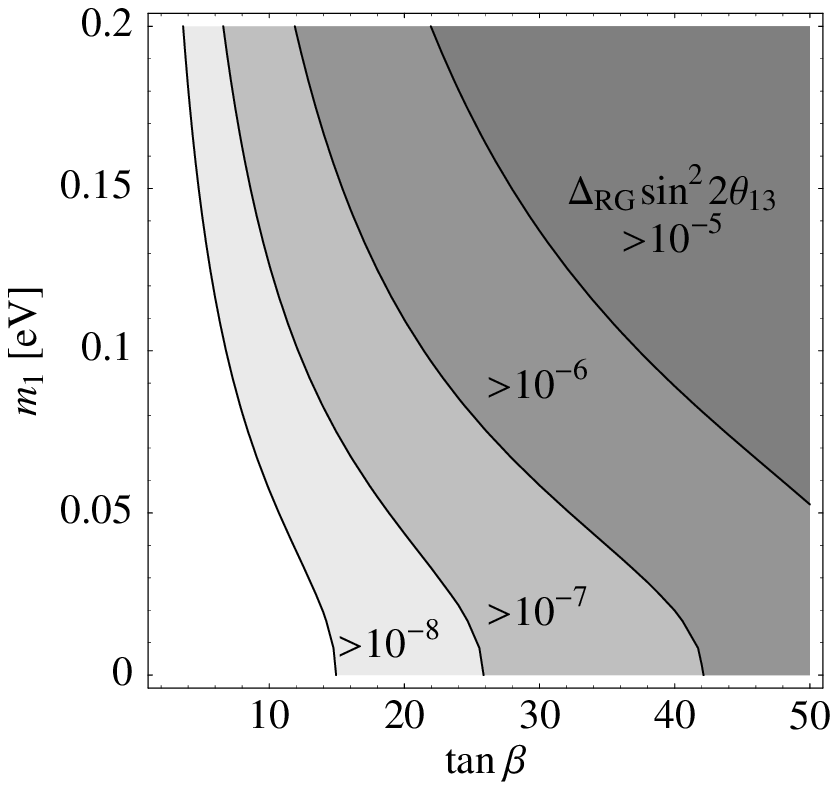}}
 }
 \caption{Radiative correction to $\sin^22\theta_{13}$, defined as 
  $\Delta_\mathrm{RG}\sin^22\theta_{13}\equiv
  |\sin^22\theta_{13}(M_\mathrm{GUT})-\sin^22\theta_{13}(m_Z)|$, as a
  function of $\tan\beta$ and $m_1$. We take $\theta_{12}(m_Z)=33^\circ$ 
  and $\theta_{23}(m_Z)=45^\circ$.}
\label{fig:DeltaRGtheta13}
\end{figure}

The running of $\theta_{13}$ depends crucially on its initial value. 
We illustrate this by plotting the radiative correction to 
$\sin^22\theta_{13}$ in Fig.~\ref{fig:DeltaRGtheta13}. Most 
important is the second term in Eq.~\eqref{eq:DotTheta13} which 
is dominant for not too small $\theta_{13}$. As a consequence we 
find that, for $\theta_{13}=0$ at the high scale, running in 
general does not generate a measurable value at the low scale. 
Only for the most optimistic sensitivities, a quite degenerate 
spectrum and large $\tan\beta$ this conclusion can be avoided. 
On the other hand, if $\theta_{13}$ is not tiny, RG effects can 
be comparable to the precision of upcoming experiments (except 
for small $\tan\beta$). 

\begin{figure}[!h]
 \centerline{%
  \subfigure[$\theta_{13}(m_Z)=10^\circ$]{%
        \includegraphics[scale=0.85]{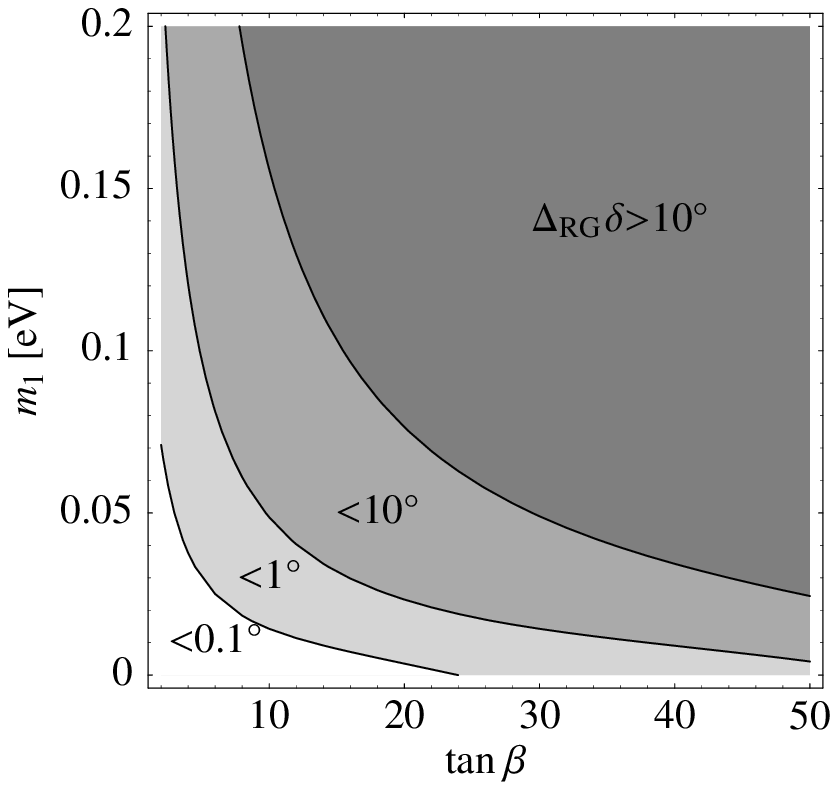}}
  \quad\subfigure[$\theta_{13}(m_Z)=0.1^\circ$]{%
        \includegraphics[scale=0.85]{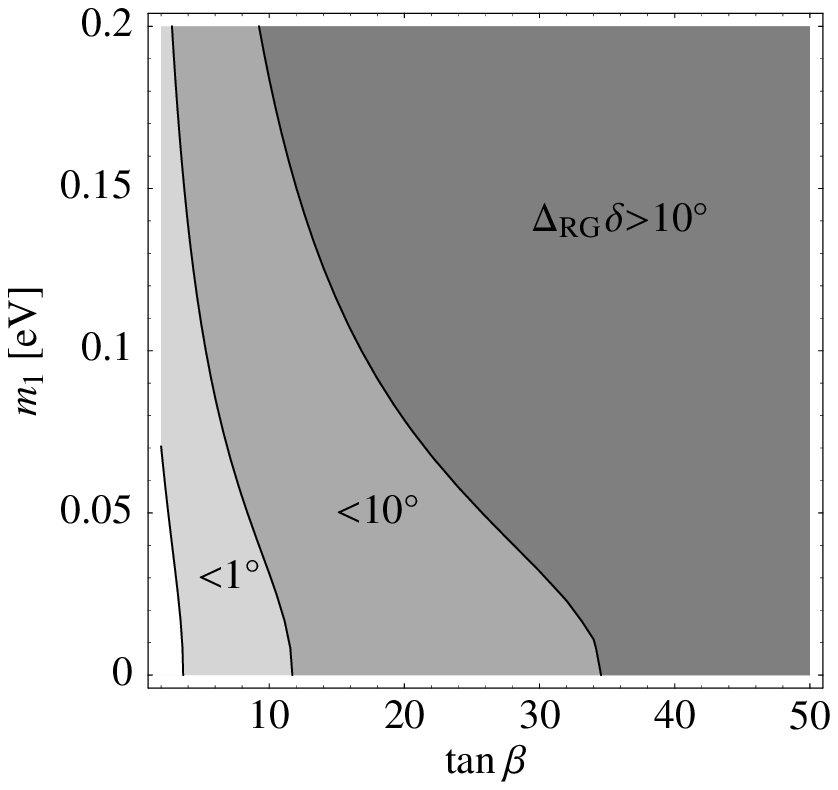}}
 }
 \caption{Radiative correction to $\delta$, defined as
  $\Delta_\mathrm{RG}\delta\equiv \big|\delta(m_Z)-
  \delta(M_\mathrm{GUT})\big|$ for (a) $\theta_{13}(m_Z)=10^\circ$ 
  and (b) $\theta_{13}(m_Z)=0.1^\circ$ as a function of $\tan\beta$ 
  and $m_1$. We use $\delta(m_Z)=90^\circ$.}
 \label{fig:DeltaRGdelta}
\end{figure}

Finally, let us discuss corrections to $\delta$. From the previous 
discussion in Sec.~\ref{sec:AnalyticFormulae} it is clear that 
small $\theta_{13}$ corresponds to an unstable configuration 
with large RG effects, even for a hierarchical spectrum 
(cf.\ Fig.~\ref{fig:DeltaRGdelta} (b)). In particular, RG effects 
are generically comparable with the precision of future experiments 
such as the combination T2K+\nova+Reactor-II, T2HK and NuFact-II (see
\cite{Huber:2004gg} and references therein).

Let us finally mention that RG effects for Dirac 
neutrinos will always result in a rescaling of the mass eigenvalues. 
Beyond that, in the framework of the SM, mixing parameters are quite 
stable. The only exceptions are $\theta_{12}$ for very degenerate 
masses, and $\delta$ for tiny $\theta_{13}$. On the other hand, in 
the MSSM, RG effects are increased by $\tan^2\beta$, i.e.\ by up to 
three orders of magnitude.

\section{Summary}

Assuming Dirac neutrinos, we have derived renormalization group 
equations for leptonic mixing parameters. The results share several 
features with the corresponding equations for Majorana neutrinos. 
However, Dirac running is more predictive, as the Majorana phases 
are unphysical in this case. This makes it possible to specify the 
amount of renormalization group evolution unambiguously as soon as 
$m_1$  and $\delta$ (and $\tan\beta$) are known. The renormalization 
group evolution alone does not yield an explanation of the largeness 
of the leptonic mixing angles (for an analogous and very clear 
discussion for Majorana neutrinos see \cite{Casas:2003kh}). Yet it 
may account for radiative enhancement of $\theta_{12}$, and possibly 
also of $\theta_{23}$, since both can increase significantly in the 
MSSM when running down.

Most importantly, we find that in phenomenological studies renormalization 
group effects for leptonic mixing angles can in general not be neglected. 
This can be understood from  the fact that 
$\Dot{\theta}_{ij} = {f(m_i,\theta_{ij},\delta)}/({m_i^2-m_j^2})$
which becomes singular if $m_i\to m_j$ and vanishes if the mixing 
angles are zero. We have thus traced back the relative enhancement 
of the quantum corrections of leptonic mixing parameters as compared 
to quark mixings to two reasons. First, the mass hierarchy 
which suppresses the renormalization group running, can be much weaker.
Second, the mixing angles are larger so that the parameters are further 
apart from the renormalization group stable situation where all mixings 
are zero.

As there is no suppression of the running by phases, the renormalization 
group corrections should in general be taken into account even for a 
strong hierarchy to accommodate the precision of future experiments. 
Renormalization group corrections are especially relevant if the 
mass spectrum is non-hierarchical, and $\tan\beta$ is large in the MSSM. 
Hence, similarly to the case of Majorana neutrinos \cite{Antusch:2003kp}, 
also in the Dirac case the non-observation of deviations of the angles 
from special points, e.g.\ of $\theta_{12}$ from
$\pi/4-\vartheta_{12}$ (with $\vartheta_{12}$ being the Cabibbo angle), 
of $\theta_{13}$ from 0 and $\theta_{23}$ from $\pi/4$, may restrict 
the parameters such as the absolute neutrino mass scale. The current 
experimental data already has the necessary precision to indicate 
disfavored parameter regions. It may also point to exactly realized symmetries 
and our formulae can hence be used to identify possible  symmetries. Whenever a
symmetry is exact and fixes some mixing parameters,  those mixing parameters
have to be stable under quantum corrections. For  instance, it has recently been
pointed out \cite{Grimus:2004cj} that  for Majorana neutrinos and an inverted
hierarchy the configuration  $m_3=\theta_{13}=0$ is stable. From the analytic
expressions it is  obvious that this statement also applies to the Dirac case.
Likewise,  a quick inspection of the RGEs excludes most of the proposed
symmetries from being exact. Our formulae are basis-independent, and thus allow
to understand certain features of the underlying theory, such as symmetries, in
a basis-independent way. We have discussed this only for the case of CP
symmetry, but it is obvious how the analysis can be carried over to other
symmetries. In this context, it would be interesting to see if infrared fixed 
points with large mixings, as discussed in  \cite{Casas:2002sn,Casas:2003kh}, 
can be obtained for (non-standard) Dirac neutrinos as well \cite{Dedes:prep}.  
In this case, one may hope for a scenario where the large mixings are a 
consequence of running, and the mechanism of generation of neutrino masses is 
still related to the scale where gauge couplings meet.

We conclude that in the light of future precision experiments, flavor 
physics might enter into an era of ``precision model building''. 
It seems possible to determine the mixing parameters to a remarkable 
accuracy, precise enough such that flavor models and the corresponding
renormalization group effects become to a certain degree distinguishable. 
For a specific parameter and a desired accuracy, our formulae allow to 
estimate the renormalization group effects, and to judge to which extent 
a numerical analysis is required.

\subsubsection*{Acknowledgements}

We would like to thank S.~Antusch, A.~Dedes and J.~Kersten for valuable
discussions. One of us (M.R.) would like to thank the Aspen Center for 
Physics for support, and the CERN theory group for hospitality. 
This work was partially supported by the EU 6th Framework Program 
MRTN-CT-2004-503369 ``Quest for Unification'' and MRTN-CT-2004-005104 
``ForcesUniverse''. This work was also supported by the Deutsche 
Forschungsgemeinschaft in the ``Sonderforschungsbereich 375  f\"ur 
Astro-Teilchenphysik'' and under project number RO-2516/3-1.

\vspace*{1cm}

\appendix
\renewcommand{\thesection}{\Alph{section}}
\renewcommand{\thesubsection}{\Alph{section}.\arabic{subsection}}
\def\theequation{\Alph{section}.\arabic{equation}}
\renewcommand{\thetable}{\arabic{table}}
\renewcommand{\thefigure}{\arabic{figure}}
\setcounter{section}{0}
\setcounter{equation}{0}

\section{Mixing parameters RGEs for Dirac masses}

\subsection{Lagrangian}

The Yukawa couplings are given by
\begin{equation}
-\mathcal{L}_\mathrm{Yuk}=\left(Y_\nu\right)_{gf}\overline{N_R^g}\tilde{\phi}^\dagger\ell_L^f+\left(Y_e\right)_{gf}\overline{e_R^g}\phi^\dagger\ell_L^f+\left(Y_u\right)_{gf}\overline{u_R^g}
 \tilde{\phi}^\dagger Q_L^f+\left(Y_d\right)_{gf}\overline{d_R^g} \phi^\dagger Q_L^f
\end{equation}
in the SM extended by right-handed neutrinos where
$\tilde{\phi}=i\sigma_2\phi^*$. In the extended MSSM, the Yukawa
couplings are analogously defined in the superpotential by 
\begin{equation}
\mathcal{W}_\mathrm{Yuk}= \left(Y_\nu\right)_{gf}
N_R^{Cg}\phi^{(2)}\epsilon^T\ell_L^f + \left(Y_e\right)_{gf} e_R^{Cg}
\phi^{(1)}\epsilon\ell_L^f + \left(Y_u\right)_{gf} u_R^{Cg}\phi^{(2)}\epsilon^T
Q_L^f+\left(Y_d\right)_{gf}d_R^{Cg}\phi^{(1)}\epsilon Q_L^f\; .
\end{equation}
The left-handed lepton and quark doublets are denoted by $\ell_L$ and $Q_L$,
respectively.
We assume that there is no Majorana mass term for the right-handed neutrinos.

\subsection{$\boldsymbol{\beta}$-functions}

The relevant $\beta$-functions for the down-type quark, up-type quark, charged
lepton and neutrino Yukawa coupling matrices $Y_d$, $Y_u$, $Y_e$ and $Y_\nu$ 
read at one-loop~\cite{Machacek:1984fi,Grzadkowski:1987tf}
\begin{subequations}
\begin{eqnarray}
 (4\pi)^2\Dot{Y}_d
 & = & 
 Y_d\left\{ 
        C_d^d\,Y_d^\dagger Y_d 
        +C_d^u\, Y_u^\dagger Y_u 
        +\alpha_d
 \right\}\;,
 \\
 (4\pi)^2 \Dot{Y}_u
 & = & Y_u\left\{
 C_u^d\,Y_d^\dagger Y_d 
 + C_u^u\, Y_u^\dagger Y_u 
 + \alpha_u
 \right\}\;,
 \\
 (4\pi)^2 \Dot{Y}_e
 & = & 
 Y_e\left\{ 
        C_e^e\,Y_e^\dagger Y_e 
        +C_e^\nu\, Y_\nu^\dagger Y_\nu 
        +\alpha_\ell
 \right\}\;,
 \\
 (4\pi)^2 \Dot{Y}_\nu
 & = & Y_\nu\left\{
 C_\nu^e\,Y_e^\dagger Y_e 
 + C_\nu^\nu\, Y_\nu^\dagger Y_\nu 
 + \alpha_\nu
 \right\}\;,\label{eq:DiracRGE}
\end{eqnarray}
\end{subequations}
where
\begin{subequations}
\begin{align}
 C_d^d 
 & = 
 \left\{\begin{array}{ll}
        3/2\;, & \text{(SM)}\\
        3\;, & \text{(MSSM)}
 \end{array}\right.
 &
 C_d^u 
 & =
 \left\{\begin{array}{ll}
        -3/2\;, & \text{(SM)}\\
        1\;, & \text{(MSSM)}
 \end{array}\right.\\
 C_u^d 
 & = 
 \left\{\begin{array}{ll}
        -3/2\;, & \text{(SM)}\\
        1\;, & \text{(MSSM)}
 \end{array}\right.
 &
 C_u^u 
 & = 
 \left\{\begin{array}{ll}
        3/2\;, & \text{(SM)}\\
        3\;, & \text{(MSSM)}
 \end{array}\right.
 \\
 C_e^e 
 & = 
 \left\{\begin{array}{ll}
        3/2\;, & \text{(SM)}\\
        3\;, & \text{(MSSM)}
 \end{array}\right.
 &
 C_e^\nu 
 & =
 \left\{\begin{array}{ll}
        -3/2\;, & \text{(SM)}\\
        1\;, & \text{(MSSM)}
 \end{array}\right.\\
 C_\nu^e 
 & = 
 \left\{\begin{array}{ll}
        -3/2\;, & \text{(SM)}\\
        1\;, & \text{(MSSM)}
 \end{array}\right.
 &
 C_\nu^\nu 
 & = 
 \left\{\begin{array}{ll}
        3/2\;, & \text{(SM)}\\
        3\;, & \text{(MSSM)}
 \end{array}\right.
\end{align}
\end{subequations}
and
\begin{subequations}
\begin{eqnarray}
 \alpha_d 
 & = &
 \left\{\begin{array}{ll}
        - \frac{1}{4} g_1^2 - \frac{9}{4} g_2^2 - 8\,g_3^2
        + T_\mathrm{SM}\;,
 & \text{(SM)}\\[0.2cm]
 3\,\Tr(Y_d^\dagger Y_d) 
 + \Tr(Y_e^\dagger Y_e)
 - \frac{7}{15}\,g_1^2
 - 3\,g_2^2 
 - \frac{16}{3}\,g_3^2\;,
 & \text{(MSSM)}
 \end{array}\right.
 \\[0.2cm]
 \alpha_u
 & = &
 \left\{\begin{array}{ll}
        - \frac{17}{20} g_1^2 - \frac{9}{4} g_2^2 - 8\,g_3^2
    + T_\mathrm{SM}\;,
 & \text{(SM)}\\[0.2cm]
 \Tr( Y_\nu ^\dagger Y_\nu ) 
 + 3\,\Tr(Y_u^\dagger Y_u)
 - \frac{13}{15}\,g_1^2 
 - 3\,g_2^2 
 - \frac{16}{3}\,g_3^2 \hphantom{- \frac{16}{3}\,g_3^2}\;,
 & \text{(MSSM)}
 \end{array}\right.
 \\
 \alpha_\ell 
 & = &
 \left\{\begin{array}{ll}
        - \frac{9}{4} g_1^2 - \frac{9}{4} g_2^2
        + T_\mathrm{SM}\;,
 & \text{(SM)}\\[0.2cm]
 3\,\Tr(Y_e^\dagger Y_e) 
 + \Tr(Y_\nu^\dagger Y_\nu)
 - \frac{9}{5}\,g_1^2
 - 3\,g_2^2 \hphantom{- \frac{16}{3}\,g_3^2}\;,
 & \text{(MSSM)}
 \end{array}\right.
 \\[0.2cm]
 \alpha_\nu
 & = &
 \left\{\begin{array}{ll}
        - \frac{9}{20}\,g_1^2 - \frac{9}{4}\,g_2^2 
    + T_\mathrm{SM}\;,
 & \text{(SM)}\\[0.2cm]
 \Tr( Y_\nu ^\dagger Y_\nu ) 
 + 3\,\Tr(Y_u^\dagger Y_u)
 - \frac{3}{5}\,g_1^2 
 - 3\,g_2^2  \;,
 & \text{(MSSM)}\;.
 \end{array}\right.
 \label{eq:AlphaNu}
\end{eqnarray} 
\end{subequations}
Here, we define $T_\mathrm{SM}\equiv\Tr\left[Y_e^\dagger Y_e + Y_\nu^\dagger
Y_\nu + 3\,Y_d^\dagger Y_d + 3\,Y_u^\dagger Y_u\right]$, and use GUT
normalization for $g_1$.

\subsection{General derivation}
\label{subsec:MixingParameterRGEsDirac}

In this subsection, we will perform a general analysis applicable for any Dirac
masses, and treat the evolution of lepton and quark masses and mixings only as a
special case.

We derive the running of mixing parameters for a RGE of the form
\begin{equation}
 16\pi^2\,\frac{\D}{\D t} H
 ~=~ 
 F^\dagger\,H+H\,F + f\, H\;,
\end{equation}
where $f$ is real and $H$ is Hermitean, so that we can diagonalize it in a
`reference basis',
\begin{equation}
 U^\dagger\cdot H\cdot U
 ~=~
 D\;.
\end{equation}
In the application in the main part, $F$ corresponds either to
$C_d^u\,Y_u^\dagger Y_u+C_d^dY_d^\dagger Y_d$ (or $C_\nu^e\,Y_e^\dagger
Y_e+C_\nu^\nu Y_\nu^\dagger Y_\nu$ for the lepton sector), and $H$ to
$Y_d^\dagger Y_d$ (or $Y_\nu^\dagger Y_\nu$). The `reference basis' is the
basis where $Y_u^\dagger Y_u$ (or $Y_e^\dagger Y_e$) is diagonal at $t=t_0$. $U$
denotes then to the CKM matrix $U_\mathrm{CKM}$ (or the MNS matrix
$U_\mathrm{MNS}$). $f$ denotes the diagonal parts of the $\beta$-function,
$f=2\alpha_d$ (or $f=2\alpha_\nu$).

Now we perform an analysis very similar to what is done in
\cite{Antusch:2003kp} which is based on  
\cite{Grzadkowski:1987tf,Babu:1987im,Casas:1999tg}. We can
differentiate the relation $H=U\cdot D\cdot U^\dagger$,
\begin{eqnarray}
 \frac{\D}{\D t}(U\cdot D\cdot U^\dagger)
 & = &
 \Dot U\cdot D\cdot U^\dagger + U\cdot D\cdot \Dot U^\dagger 
        + U\cdot \Dot D\cdot U^\dagger
 \nonumber\\ 
 & \stackrel{!}{=} &
 \frac{1}{16\pi^2}      \left(F^\dagger \cdot U \cdot D \cdot U^\dagger
        +U\cdot D\cdot U^\dagger \cdot F + f\,U\cdot D \cdot U^\dagger\right)
 \;.            
\end{eqnarray}
Multiplying by $U^\dagger$ from the left and by $U$ from the right yields
\begin{equation}
 U^\dagger\cdot \Dot U \cdot D + D\cdot \Dot U^\dagger\cdot U + \Dot D
 ~=~\frac{1}{16\pi^2}
 \left(F^{\prime\,\dagger}\cdot D + D\cdot F' +f\,D\right)
 \;,
\end{equation}
where $F'=U^\dagger\cdot F\cdot U$. For the quark case,
$F'=C_d^d\,D+C_d^u\,U^\dagger Y_u^\dagger Y_u\,U$. We will see below that only
the off-diagonal components are relevant for the RGEs of the mixing parameters.

The evolution of $U$ can be written as
\begin{equation}\label{eq:EvolutionOfU}
 \frac{\D}{\D t} U~=~U\cdot X\;,
\end{equation}
where $X$ is anti-Hermitean.
Inserting this relation yields
\begin{equation} 
 \Dot D +X\cdot D + D\cdot X^\dagger
 ~=~
 \frac{1}{16\pi^2}\left(F^{\prime\,\dagger}\cdot D + D\cdot F' +f\,D\right)
 \;,
\end{equation}
or, by using the anti-Hermitecity of $X$,
\begin{equation}
 \Dot D
 ~=~
 \frac{1}{16\pi^2}\left(f\,D +F^{\prime\,\dagger}\cdot D + D\cdot F'\right)
 -X\cdot D + D\cdot X\;.
\end{equation}
Denoting the entries of $D$ by $y_i^2$, i.e.\ $D=\diag(y_1^2,y_2^2,y_3^2)$, we
find
\begin{equation}
 \frac{\D}{\D t}y_i^2
 ~=~
 \frac{1}{16\pi^2}\left[f\,y_i^2+(F_{ii}^{\prime\,*}+F_{ii}')\,y_i^2\right]
 \;,
\end{equation}
i.e.\ the terms proportional to $X$ have dropped out. This equation corresponds
to a RGE for the running mass eigenvalues, defined by $m_i(t)=|y_i(t)|\,v$ with
$v$ fixed, of the form
\begin{equation}
 (4\pi)^2\Dot{m}_i
 ~=~
 (\re F_{ii}'+\alpha)\,m_i\;.
\end{equation}
By analyzing the off-diagonal parts we obtain
\begin{equation}
 y_i^2\,X_{ij} - X_{ij}\,y_j^2
 ~=~
 -\frac{1}{16\pi^2}\left[(F^{\prime\,\dagger})_{ij}\,y_j^2
 +y_i^2\,F'_{ij}\right]
 \;.
\end{equation}
This can be converted into equations for real and imaginary part of $X$,
which, since $F$ is Hermitean, can be combined to
\begin{equation} \label{eq:XfromFprime}
 16\pi^2 X_{ij} = \frac{y_j^2+y_i^2}{y_j^2-y_i^2} \, F_{ij}' \;.
\end{equation}

The diagonal parts of $X$ remain undetermined.  However, this is not a problem,
since they only influence the RG evolution of the unphysical
phases.\footnote{Note that the Majorana phases are unphysical in the the Dirac
case as well.}

So far, we have derived the differential change of the CKM matrix due to the RG
corrections for $Y_d^\dagger Y_d$ (cf.\ Eq.~\eqref{eq:EvolutionOfU}). In the
Majorana neutrino case, the analogous differential equation already describes
the evolution of the MNS matrix since $Y_e^\dagger Y_e$ doesn't get rotated by
the RGE.\footnote{This is only true at leading order in $M^{-1}$ where $M$ 
denotes the scale of the effective neutrino mass operator (e.g.\ the see-saw
scale) \cite{Broncano:2004tz}.} For Dirac neutrinos, $Y_e^\dagger Y_e$ gets
rotated only by terms proportional to the squares of Dirac Yukawa couplings,
hence those rotations can safely be neglected. In the quark sector, the
radiative  rotation of $Y_u^\dagger Y_u$ represents an important effect, as we
will argue in the following.

\subsection{Contribution from the change of $\boldsymbol{Y_u}$}

Here, we specialize to the quark sector as the analogous effect is irrelevant
for Dirac neutrinos.

The RGE for $Y_u$ contains non-diagonal terms so that continuous
re-diagonalization is required. Since the mixing matrix $U_\mathrm{CKM}$ is
defined as the matrix which  diagonalizes $Y_d^\dagger Y_d$ in the basis in
which $Y_u$ is diagonal, $U_\mathrm{CKM}$ receives an additional contribution
from the running of $Y_u$,
\begin{equation}
 \frac{\D}{\D t}U_\mathrm{CKM}
 ~=~
 U_\mathrm{CKM}\cdot X+ \text{term stemming from the change of}\:Y_u\;.
\end{equation}

To evaluate this change, we can essentially repeat the steps of the previous
subsection. Introducing a matrix $\widetilde{U}$ which diagonalizes $Y_u^\dagger
Y_u$ in the reference basis (implying $\widetilde{U}(t=t_0)=\mathbbm{1}$), i.e.\
\begin{equation}
 \widetilde{U}^\dagger\,Y_u^\dagger Y_u\,\widetilde{U}
 ~=~
 \diag(\widetilde{y}_1^2,\widetilde{y}_2^2,\widetilde{y}_3^2)\;,
\end{equation}
we arrive at
\begin{equation}
 \frac{\D}{\D t}\widetilde{U}
 ~=~
 \widetilde{U}\cdot \widetilde{X}\;,
\end{equation}
where the off-diagonal entries of $\widetilde{X}$ are given by
\begin{equation}\label{eq:Xtilde}
 16\pi^2\,\widetilde{X}_{ij}
 ~=~
 \frac{\widetilde{y}_i^2+\widetilde{y}_j^2}{\widetilde{y}_j^2-\widetilde{y}_i^2}
 \,\widetilde{F}_{ij}\;.
\end{equation}
Completely analogous to \ref{subsec:MixingParameterRGEsDirac},
\begin{equation}
 \widetilde{F}'
 ~=~
 \widetilde{U}^\dagger\cdot\widetilde{F}\cdot\widetilde{U}\;,
\end{equation}
and at $t=t_0$
\begin{equation}
 \widetilde{F}'
 ~=~C_u^d\,U\,D\,U^\dagger+C_u^u\,Y_u^\dagger Y_u\;.
\end{equation}
Again, only the off-diagonal terms influence the RGEs of the mixing angles.

\subsection{Mixing parameter RGEs in the quark sector}

As $U_\mathrm{CKM}=\widetilde{U}^{-1}U=\widetilde{U}^\dagger U$, the RGE for the
CKM matrix reads
\begin{equation}
 \frac{\D}{\D t}U_\mathrm{CKM}
 ~=~
 \widetilde{X}^\dagger\cdot U_\mathrm{CKM}+U_\mathrm{CKM}\cdot X\;.
\end{equation}

To proceed, we label the mixing parameters by
\begin{equation}
 \{\xi_k\} 
 = 
 \{\theta_{12},\theta_{13},\theta_{23},
        \delta,\delta_e,\delta_\mu,\delta_\tau,
        \varphi_1,\varphi_2\}
 \;,    
\end{equation}
and evaluate the derivative of $U_\mathrm{CKM}$,
\begin{equation}
 \Dot U_\mathrm{CKM}
 ~=~
 \Dot U_\mathrm{CKM} \left(\{\Dot \xi_k\},\left\{\xi_k\right\}\right)
 \;.
\end{equation}
Observe that the resulting expression is linear in $\Dot\xi_k$. By solving a
system of linear equations of the form
\begin{equation}
 \sum\limits_k
  A^{(k)}_{TX}\,\Dot\xi_k + \I\,S^{(k)}_{TX}\,\Dot\xi_k
 ~=~
 R_{X}\;,
\end{equation}
where
\begin{equation}
 R_{X}
 ~=~
 U_\mathrm{MNS}\cdot T + X^\dagger \cdot  U_\mathrm{MNS}\;,
\end{equation}
we thus obtain a set of linear equations for the $\Dot\xi_k$.
RGEs for the matrix elements have been derived in refs.~\cite{Babu:1987im,Grzadkowski:1987tf}.

From these, we obtain the RGEs for the mixing angles in the quark sector.
Neglecting all Yukawa coupling except for $y_t$ and $y_b$, they read
\begin{subequations}
\begin{eqnarray}
 \Dot{\vartheta}_{12}
 & = &
 \frac{C_{d}^{u}\,{{y_t}}^2}{64\,{\pi }^2}
 \,\cos ({{\vartheta }_{12}})\,
    \left\{ \left[
                \left( 3 - \cos 2\,\vartheta_{13} \right) \,\cos 2\,\vartheta_{23}
                - 2\,\cos^2\vartheta _{13} \right] \,\sin \vartheta_{12} 
 \right.\nonumber\\
 & & \qquad\left.{}
 + 4\,\cos \delta_\mathrm{CP} \,\cos \vartheta_{12}\,
       \sin \vartheta_{13}\,\sin 2\,\vartheta _{23} \right\}
 \;,\\
 \Dot{\vartheta}_{13}
 & = &
 \frac{-\sin 2\,\vartheta_{13}}{64\,{\pi }^2}
 \left[ 2\,C_{u}^{d}\,y_b^2 + C_{d}^{u}\,y_t^2\, 
 	\left(1+ \cos 2\,\vartheta_{23}\right) \right] 
 \;,\\
 \Dot{\vartheta}_{23}
 & = &
 \frac{-\sin 2\,\vartheta_{23}}{64\,{\pi }^2}
 \left[ C_{u}^{d}\,y_b^2\,\left(1+\cos 2\,\vartheta_{13}\right) 
 	+ 2\,C_{d}^{u}\,y_t^2  \right] 
 \;.
\end{eqnarray}
\end{subequations}
It turns out that finite $y_s$ and $y_c$ corrections yield an important but
sub-dominant effect for $\Dot{\vartheta}_{12}$.  The dominant term in the RGE of
$\delta_\mathrm{CP}$ is
\begin{equation}
 \Dot{\delta}_\mathrm{CP}~=~
 \frac{C_{d}^{u}\,y_s^2\,y_t^2}{8\,\pi^2\,
    \left( y_b^2 - y_s^2 \right) }
 \,\cos \vartheta_{12}\,\cos \vartheta_{23}\,\sin\delta\,\sin \vartheta_{12}\,
    \sin \vartheta_{23}\times\vartheta_{13}^{-1}\;.
\end{equation}

\subsection{Mixing parameter RGEs in the (Dirac) neutrino sector}

In order to derive analogous RGEs for the leptonic mixing parameters, observe
that the RG change of $Y_e^\dagger Y_e$ is quadratic in
neutrino Yukawa couplings, i.e.\ strongly suppressed. Thus, we can safely
neglect the $\widetilde{X}$ contribution,
\begin{equation}
 \frac{\D}{\D t}U_\mathrm{MNS}
 ~=~
 \widetilde{X}^\dagger\cdot U_\mathrm{MNS}+U_\mathrm{MNS}\cdot X
 ~\simeq~
 U_\mathrm{MNS}\cdot X\;,
\end{equation}
where $X$ is now related to $F'$ by Eq.~\eqref{eq:XfromFprime}, and
$F'=C_\nu^\nu\,D+C_\nu^e\,U_\mathrm{MNS}^\dagger Y_e^\dagger
Y_e\,U_\mathrm{MNS}$ at $t=t_0$.

\bibliography{Running}

\providecommand{\bysame}{\leavevmode\hbox to3em{\hrulefill}\thinspace}
\begin{thebibliography}{10}

\bibitem{Minkowski:1977sc}
P.~Minkowski, Phys. Lett. \textbf{B67} (1977), 421.

\bibitem{Ramond:1979py}
P.~Ramond, \emph{The family group in grand unified theories}, 1979, Invited
  talk given at Sanibel Symposium, Palm Coast, Fla., Feb 25 - Mar 2, 1979.

\bibitem{Yanagida:1980}
T.~Yanagida, \emph{Horizontal gauge symmetry and masses of neutrinos}, in
  \emph{Proceedings of the Workshop on The Unified Theory and the Baryon Number
  in the Universe} (O.~Sawada and A.~Sugamoto, eds.), KEK, Tsukuba, Japan,
  1979, p.~95.

\bibitem{Glashow:1979vf}
S.~L. Glashow, \emph{The future of elementary particle physics}, in
  \emph{Proceedings of the 1979 Carg{\`e}se Summer Institute on Quarks and
  Leptons} (M.~L{\'e}vy, J.-L. Basdevant, D.~Speiser, J.~Weyers, R.~Gastmans,
  and M.~Jacob, eds.), Plenum Press, New York, 1980, pp.~687--713.

\bibitem{Gell-Mann:1980vs}
M.~Gell-Mann, P.~Ramond, and R.~Slansky, \emph{Complex spinors and unified
  theories}, in \emph{Supergravity} (P.~van Nieuwenhuizen and D.~Z. Freedman,
  eds.), North Holland, Amsterdam, 1979, p.~315.

\bibitem{Mohapatra:1980ia}
R.~N. Mohapatra and G.~Senjanovi{\'c}, Phys. Rev. Lett. \textbf{44} (1980),
  912.

\bibitem{Mohapatra:1986bd}
R.~N. Mohapatra and J.~W.~F. Valle, Phys. Rev. \textbf{D34} (1986), 1642.

\bibitem{Arkani-Hamed:2000bq}
N.~Arkani-Hamed, L.~J. Hall, H.~Murayama, D.~R. Smith, and N.~Weiner, Phys.
  Rev. \textbf{D64} (2001), 115011,  [hep-ph/0006312].

\bibitem{Borzumati:2000mc}
F.~Borzumati and Y.~Nomura, Phys. Rev. \textbf{D64} (2001), 053005,
  [hep-ph/0007018].

\bibitem{Kitano:2002px}
R.~Kitano, Phys. Lett. \textbf{B539} (2002), 102--106,  [hep-ph/0204164].

\bibitem{Arnowitt:2003kc}
R.~Arnowitt, B.~Dutta, and B.~Hu, Nucl. Phys. \textbf{B682} (2004), 347--366,
  [hep-th/0309033].

\bibitem{Abel:2004tt}
S.~Abel, A.~Dedes, and K.~Tamvakis, Phys. Rev. \textbf{D71} (2005), 033003,
  [hep-ph/0402287].

\bibitem{Hung:2004ac}
P.~Q. Hung, Nucl. Phys. \textbf{B720} (2005), 89--115,  [hep-ph/0412262].

\bibitem{Ko:2005sh}
P.~Ko, T.~Kobayashi, and J.-H. Park, Phys. Rev. \textbf{D71} (2005), 095010,
  [hep-ph/0503029].

\bibitem{Antusch:2005kf}
S.~Antusch, O.~J. Eyton-Williams, and S.~F. King,  (2005),  hep-ph/0505140.

\bibitem{Giedt:2005vx}
J.~Giedt, G.~L. Kane, P.~Langacker, and B.~D. Nelson,  (2005),  hep-th/0502032.

\bibitem{Murayama:2004me}
H.~Murayama, Nucl. Phys. Proc. Suppl. \textbf{137} (2004), 206--219,
  [hep-ph/0410140].

\bibitem{Smirnov:2004hs}
A.~Y. Smirnov,  (2004),  hep-ph/0411194.

\bibitem{Mohapatra:2004vr}
R.~N. Mohapatra et~al.,  (2004),  hep-ph/0412099.

\bibitem{Fukugita:1986hr}
M.~Fukugita and T.~Yanagida, Phys. Lett. \textbf{174B} (1986), 45.

\bibitem{Buchmuller:2005eh}
W.~Buchm{\"u}ller, R.~D. Peccei, and T.~Yanagida,  (2005),  hep-ph/0502169.

\bibitem{Dick:1999je}
K.~Dick, M.~Lindner, M.~Ratz, and D.~Wright, Phys. Rev. Lett. \textbf{84}
  (2000), 4039--4042,  [hep-ph/9907562].

\bibitem{Murayama:2002je}
H.~Murayama and A.~Pierce, Phys. Rev. Lett. \textbf{89} (2002), 271601,
  [hep-ph/0206177].

\bibitem{Dolgov:1994vu}
A.~D. Dolgov, K.~Kainulainen, and I.~Z. Rothstein, Phys. Rev. \textbf{D51}
  (1995), 4129--4133,  [hep-ph/9407395].

\bibitem{Chiang:2000um}
C.-W. Chiang, Phys. Rev. \textbf{D63} (2001), 076009,  [hep-ph/0011195].

\bibitem{Eidelman:2004wy}
Particle Data Group, S.~Eidelman et~al., Phys. Lett. \textbf{B592} (2004), 1.

\bibitem{Antusch:2003kp}
S.~Antusch, J.~Kersten, M.~Lindner, and M.~Ratz, Nucl. Phys. \textbf{B674}
  (2003), 401--433,  [hep-ph/0305273].

\bibitem{Antusch:2001ck}
S.~Antusch, M.~Drees, J.~Kersten, M.~Lindner, and M.~Ratz, Phys. Lett.
  \textbf{B519} (2001), 238--242,  [hep-ph/0108005].

\bibitem{Antusch:2001vn}
S.~Antusch, M.~Drees, J.~Kersten, M.~Lindner, and M.~Ratz, Phys. Lett.
  \textbf{B525} (2002), 130--134,  [hep-ph/0110366].

\bibitem{Chankowski:1993tx}
P.~H. Chankowski and Z.~Pluciennik, Phys. Lett. \textbf{B316} (1993), 312--317,
   [hep-ph/9306333].

\bibitem{Babu:1993qv}
K.~S. Babu, C.~N. Leung, and J.~Pantaleone, Phys. Lett. \textbf{B319} (1993),
  191--198,  [hep-ph/9309223].

\bibitem{Casas:1999tg}
J.~A. Casas, J.~R. Espinosa, A.~Ibarra, and I.~Navarro, Nucl. Phys.
  \textbf{B573} (2000), 652,  [hep-ph/9910420].

\bibitem{Antusch:2005gp}
S.~Antusch, J.~Kersten, M.~Lindner, M.~Ratz, and M.~A. Schmidt, JHEP
  \textbf{03} (2005), 024,  [hep-ph/0501272].

\bibitem{Suzuki:2004pb}
Y.~Suzuki, Nucl. Phys. Proc. Suppl. \textbf{137} (2004), 5--14.

\bibitem{Gonzalez-Garcia:2004jd}
M.~C. Gonzalez-Garcia,  (2004),  hep-ph/0410030.

\bibitem{Fukuda:2002pe}
Super-Kamiokande, S.~Fukuda et~al., Phys. Lett. \textbf{B539} (2002), 179--187,
   [hep-ex/0205075].

\bibitem{Cleveland:1998nv}
B.~T. Cleveland et~al., Astrophys. J. \textbf{496} (1998), 505--526.

\bibitem{Abdurashitov:1999zd}
SAGE, J.~N. Abdurashitov et~al., Phys. Rev. \textbf{C60} (1999), 055801,
  [astro-ph/9907113].

\bibitem{Hampel:1998xg}
GALLEX, W.~Hampel et~al., Phys. Lett. \textbf{B447} (1999), 127--133.

\bibitem{Cattadori:2002rd}
GNO, C.~M. Cattadori, Nucl. Phys. Proc. Suppl. \textbf{110} (2002), 311--314.

\bibitem{Ahmad:2002jz}
SNO Collaboration, Q.~R. Ahmad et~al., Phys. Rev. Lett. \textbf{89} (2002),
  011301,  [nucl-ex/0204008].

\bibitem{Ahmad:2002ka}
SNO Collaboration, Q.~R. Ahmad et~al.,  (2002),  nucl-ex/0204009.

\bibitem{Ahmed:2003kj}
SNO, S.~N. Ahmed et~al., Phys. Rev. Lett. \textbf{92} (2004), 181301,
  [nucl-ex/0309004].

\bibitem{Suzuki:2004pd}
KamLAND, A.~Suzuki, Nucl. Phys. Proc. Suppl. \textbf{137} (2004), 21--27.

\bibitem{Apollonio:1997xe}
CHOOZ, M.~Apollonio et~al., Phys. Lett. \textbf{B420} (1998), 397--404,
  [hep-ex/9711002].

\bibitem{Apollonio:2002gd}
M.~Apollonio et~al., Eur. Phys. J. \textbf{C27} (2003), 331--374,
  [hep-ex/0301017].

\bibitem{Boehm:1999gk}
F.~Boehm et~al., Phys. Rev. Lett. \textbf{84} (2000), 3764--3767,
  [hep-ex/9912050].

\bibitem{Maltoni:2004ei}
M.~Maltoni, T.~Schwetz, M.~A. Tortola, and J.~W.~F. Valle, New J. Phys.
  \textbf{6} (2004), 122,  [hep-ph/0405172].

\bibitem{Maltoni:2003da}
M.~Maltoni, T.~Schwetz, M.~A. Tortola, and J.~W.~F. Valle, Phys. Rev.
  \textbf{D68} (2003), 113010,  [hep-ph/0309130].

\bibitem{Freund:2000ti}
M.~Freund, P.~Huber, and M.~Lindner, Nucl. Phys. \textbf{B585} (2000),
  105--123,  [hep-ph/0004085].

\bibitem{Freund:2001ui}
M.~Freund, P.~Huber, and M.~Lindner,  (2001),  hep-ph/0105071.

\bibitem{Huber:2002rs}
P.~Huber, M.~Lindner, and W.~Winter, Nucl. Phys. \textbf{B654} (2003), 3--29,
  [hep-ph/0211300].

\bibitem{Lindner:2002vt}
M.~Lindner, Springer Tracts Mod. Phys. \textbf{190} (2003), 209--242,
  [hep-ph/0209083].

\bibitem{Huber:2002mx}
P.~Huber, M.~Lindner, and W.~Winter, Nucl. Phys. \textbf{B645} (2002), 3--48,
  [hep-ph/0204352].

\bibitem{Ables:1995wq}
MINOS, E.~Ables et~al., FERMILAB-PROPOSAL-0875.

\bibitem{Aprili:2002wx}
ICARUS, P.~Aprili et~al., CERN-SPSC-2002-027.

\bibitem{Duchesneau:2002yq}
OPERA, D.~Duchesneau, eConf \textbf{C0209101} (2002), TH09,  [hep-ex/0209082].

\bibitem{Itow:2001ee}
Y.~Itow et~al., \emph{The {JHF-Kamioka} neutrino project}, in \emph{Proceedings
  of the 3rd Workshop on Neutrino Oscillations and their Origin (NOON 2001)}
  (Y.~Suzuki, M.~Nakahata, Y.~Fukuda, Y.~Takeuchi, T.~Mori, and T.~Yoshida,
  eds.), World Scientific, Singapore, 2003, p.~239.

\bibitem{Ayres:2002nm}
Nova, D.~Ayres et~al.,  (2002),  hep-ex/0210005.

\bibitem{Anderson:2004pk}
K.~Anderson et~al.,  (2004),  hep-ex/0402041.

\bibitem{Huber:2004dv}
P.~Huber, M.~Lindner, M.~Rolinec, T.~Schwetz, and W.~Winter,  (2004),
  hep-ph/0412133.

\bibitem{Minakata:2004pg}
H.~Minakata, M.~Sonoyama, and H.~Sugiyama, Phys. Rev. \textbf{D70} (2004),
  113012,  [hep-ph/0406073].

\bibitem{Huber:2003pm}
P.~Huber, M.~Lindner, T.~Schwetz, and W.~Winter,  (2003),  hep-ph/0303232.

\bibitem{Choubey:2004bf}
S.~Choubey and S.~T. Petcov, Phys. Lett. \textbf{B594} (2004), 333--346,
  [hep-ph/0404103].

\bibitem{Huber:2005jk}
P.~Huber, M.~Lindner, M.~Rolinec, and W.~Winter,  (2005),  hep-ph/0506237.

\bibitem{Gogoladze:2001kj}
I.~Gogoladze and A.~Perez-Lorenzana, Phys. Rev. \textbf{D65} (2002), 095011,
  [hep-ph/0112034].

\bibitem{Hagedorn:2005kz}
C.~Hagedorn and W.~Rodejohann, JHEP \textbf{07} (2005), 034,  [hep-ph/0503143].

\bibitem{Raidal:2004iw}
M.~Raidal, Phys. Rev. Lett. \textbf{93} (2004), 161801,  [hep-ph/0404046].

\bibitem{Minakata:2004xt}
H.~Minakata and A.~Y. Smirnov, Phys. Rev. \textbf{D70} (2004), 073009,
  [hep-ph/0405088].

\bibitem{Frampton:2004vw}
P.~H. Frampton and R.~N. Mohapatra, JHEP \textbf{01} (2005), 025,
  [hep-ph/0407139].

\bibitem{Mei:2005qp}
J.-W. Mei, Phys. Rev. \textbf{D71} (2005), 073012,  [hep-ph/0502015].

\bibitem{Ellis:2005dr}
J.~Ellis, A.~Hektor, M.~Kadastik, K.~Kannike, and M.~Raidal,  (2005),
  hep-ph/0506122.

\bibitem{Huber:2004gg}
P.~Huber, M.~Lindner, and W.~Winter,  (2004),  hep-ph/0412199.

\bibitem{Casas:2003kh}
J.~A. Casas, J.~R. Espinosa, and I.~Navarro, JHEP \textbf{09} (2003), 048,
  [hep-ph/0306243].

\bibitem{Grimus:2004cj}
W.~Grimus and L.~Lavoura, J. Phys. \textbf{G31} (2005), 683--692,
  [hep-ph/0410279].

\bibitem{Casas:2002sn}
J.~A. Casas, J.~R. Espinosa, and I.~Navarro, Phys. Rev. Lett. \textbf{89}
  (2002), 161801,  [hep-ph/0206276].

\bibitem{Dedes:prep}
A.~Dedes, in preparation.

\bibitem{Machacek:1984fi}
M.~E. Machacek and M.~T. Vaughn, Nucl. Phys. \textbf{B236} (1984), 221.

\bibitem{Grzadkowski:1987tf}
B.~Grzadkowski and M.~Lindner, Phys. Lett. \textbf{B193} (1987), 71.

\bibitem{Babu:1987im}
K.~S. Babu, Z. Phys. \textbf{C35} (1987), 69.

\bibitem{Broncano:2004tz}
A.~Broncano, M.~B. Gavela, and E.~Jenkins, Nucl. Phys. \textbf{B705} (2005),
  269--295,  [hep-ph/0406019].

\end{thebibliography}
\bibliographystyle{ArXiv}

\end{document}